# Plasma–wall interaction in laser inertial fusion reactors: novel proposals for radiation tests of first wall materials

J Alvarez Ruiz , A Rivera , K Mima , D Garoz , R Gonzalez-Arrabal ,
N Gordillo , J Fuchs , K Tanaka , I Fernández , F Briones  and
J Perlado

**Abstract**
Dry-wall laser inertial fusion (LIF) chambers will have to withstand strong bursts of fast charged particles which will deposit tens of kJ m$^{-2}$ and implant more than $10^{18}$ particles m$^{-2}$ in a few microseconds at a repetition rate of some Hz. Large chamber dimensions and resistant plasma-facing materials must be combined to guarantee the chamber performance as long as possible under the expected threats: heating, fatigue, cracking, formation of defects, retention of light species, swelling and erosion. Current and novel radiation resistant materials for the first wall need to be validated under realistic conditions. However, at present there is a lack of facilities which can reproduce such ion environments.

This contribution proposes the use of ultra-intense lasers and high-intense pulsed ion beams (HIPIB) to recreate the plasma conditions in LIF reactors. By target normal sheath acceleration, ultra-intense lasers can generate very short and energetic ion pulses with a spectral distribution similar to that of the inertial fusion ion bursts, suitable to validate fusion materials and to investigate the barely known propagation of those bursts through background plasmas/gases present in the reactor chamber. HIPIB technologies, initially developed for inertial fusion driver systems, provide huge intensity pulses which meet the irradiation conditions expected in the first wall of LIF chambers and thus can be used for the validation of materials too.

## 1. Introduction

After the construction of NIF, the National Ignition Facility at the Lawrence Livermore National Laboratory [1] and its predictions that it will obtain ignition and gain in the next few years, laser inertial fusion (LIF) stands out as a promising energy source. In short, LIF aims to reaching Lawson conditions by a sudden compression and ignition of an encapsulated mixture of D–T by focusing tens of very intense laser pulses on a target a few millimeters in diameter.

One of the main concerns in the development of laser fusion technology is the effect of the ion bursts created in the target fusion on the inner components of the reactor. Those bursts of very energetic particles coming from the ablation of the capsule and the fusion reaction itself (i.e. H, D, T, He, C and other high $Z$ materials, all of which from now on will be collectively described as fusion ions) are known to severely damage the materials of the first wall. How those ions interact with the plasma-facing components and how such an interaction reduces the operational lifetime of those elements are key to selecting the constituent materials and to designing the dimensions of the chamber. The spectral energy distribution (particles MeV$^{-1}$) of each species is crucial to evaluate their corresponding temporal and spatial energy deposition on walls. Although there are some estimations of the spectral distribution for each ion after the explosion (which are used to evaluate the ion-first wall interaction in LIF) very little is known about the influence of the surrounding gases/plasmas of the chamber on the final energy of ions. Both, the ion–matter interaction and the ion transport through a background plasma have been little investigated, mainly due to the lack of available facilities which can recreate the properties of the fusion ion pulses. However, the authors have realized the potential of ultra-intense laser systems (in particular, those of table-top size and high repetition rate) and the high-intense pulsed ion beams (HIPIB) technique as excellent tools to reproduce the main characteristics of those ion bursts and thus, tackle these questions.

This work will first describe the main properties of typical LIF ion bursts, their effects on the plasma-facing components and the importance of their final energy spectrum after their propagation through the existing plasma in the chamber. Then, the mechanism to produce ion pulses similar to those of fusion using ultra-intense lasers will be introduced, together with a simulation comparing the thermo-mechanical effects produced in both cases on a tungsten wall. Studies of laser-driven ion beams through controlled plasmas can also help to understand the interaction of fusion ions with the surrounding plasmas in the chamber, in particular the two stream instability which can modify the final energy distribution of ions. The possibility of producing HIPIB in a very simple compact and versatile magnetron-based setup will be presented giving the first experimental parameters of the generated beams. As a conclusion, we will discuss the strong impact that the proposed techniques could have on the evaluation of fusion materials and the design of laser fusion reactors.

## 2. LIF ions

In LIF, there are different ignition schemes proposed for a more energetically efficient combustion of the D–T fuel, namely, central ignition, fast ignition and the more recent shock ignition scheme. Each of them have their particular stages of compression and burn-out but the main difference in the energy spectra of the generated ion bursts stems from the coupling of the laser energy to the target. The straightforward mechanism in which the laser pulses are directly focused on the fuel pellet is called direct drive and typically produces an ion spectrum which carries away 25% of the fusion energy. In the second investigated mechanism, called indirect drive, the laser pulses are focused on the inner walls of a capsule that contains the pellet, converting first the laser energy into x-rays which later heat and compress the fuel. This indirect process generates an ion spectrum which accounts for only 1–2% of the total energy. For investigations of fusion ions-first wall interactions, ion bursts from direct drive targets are more interesting and will be the case we study throughout this work. However, all the following discussions are also valid for indirect drive targets, the ion propagation through background gases/plasmas being particularly relevant (indirect drive targets require a denser background gas as first wall protection from x-rays).

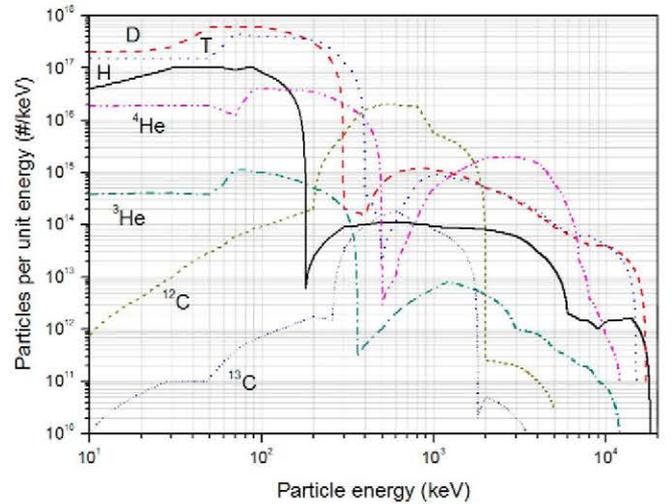

**Figure 1.** Energy spectra of LIF products for a direct drive shock ignition target of 48 MJ with permission of Dr J Perkins [4].

**Table 1.** Summary of main ion parameters of a direct drive shock ignition fusion target of 48 MJ [4]. Column one corresponds to the most abundant ions. Column two indicates the percentage of the total generated fusion energy carried away by each ion. Column three represents the average kinetic energy of each ion. Column four shows the relative quantity of each of the indicated ions.

| Particle | Total Fusion Energy (%) | Particle average energy (keV) | Total number of ions (%) |
|---|---|---|---|
| H | 0.5 | 143 | 5.1 |
| D | 6.6 | 191 | 42.2 |
| T | 7.4 | 235 | 39.7 |
| $^4$He | 7.6 | 1334 | 7.2 |
| $^{12}$C | 3.5 | 760 | 6.0 |

Figure 1 represents a typical spectral energy distribution of the different ions of a direct drive shock ignition target. Other examples of direct drive targets with a central ignition scheme can be found at the ARIES program web page [2]. In all cases, those results indicate the initial (100 ns after implosion) energy status of the different ionic species and have been calculated using the LASNEX code from LLNL [3].

The general information one can extract from those spectra is summarized in table 1. In short, direct drive targets produce ion bursts of $10^{19}$–$10^{20}$ particles per species with a high and broad energy distribution. When these ions reach the plasma-facing components, typically situated a few meters from the explosion, they do it as a high flux beam, $10^{24}$ m$^{-2}$ s$^{-1}$, with a high energy intensity of GW m$^{-2}$. As a rough estimate, ions deposit most of their energy in the first few micrometers of the components during a few microseconds, at least in the case of a high Z material wall [5].

Here, it is very important to stress the relevance of the spectral energy distribution of the ion bursts on the interaction with the walls. From the thermo-mechanical point of view, that energy distribution determines the spatial and temporal deposition profile of energy and thus, the maximum temperature and mechanical stress which the wall is exposed to [5]. In general, that interaction implies a significant increase in the temperature of the first micrometers and a stress leading to plastification and permanent deformation of the material. The repetitive nature of LIF reactors induces thermo-mechanical cycles which eventually cause fatigue and fracture of the material and the consequent reduction of operational lifetime [6]. The determination of that operational limit is of utmost importance for the design of any fusion device. From an atomistic point of view, those energetic ion bursts and their energy spectral distribution have an even more drastic effect on the first wall. Energetic ions disrupt the lattice structure of the material, removing layers by physical or chemical sputtering and generating defects. The quantity of sputtered atoms and the number and location of defects depends very much on the incident flux and energy of the ions. Each shot creates Frenkel pairs which diffuse and aggregate, producing dislocations, voids, etc. Moreover, the implanted light ions can nucleate, giving rise to highly pressurized nano-bubbles which, in turn, induce swelling, exfoliation of the material and mechanical failure [7]. Added to those effects, the simultaneous arrival of different ions cause collective processes which are barely known. Probably, from all previous effects, this radiation-induced damage by synergistic interactions are the least investigated and, in the long run, might become decisive in the design of a reactor.

From the previous discussion, it is clear that any component selected for the first wall of a nuclear fusion reactor should have a thorough validation under such an ion environment. However, there are very little studies on this respect and only some investigations carried out at RHEPP I at the Sandia National Laboratory [8] or by the inertial electrostatic confinement device at the University of Wisconsin-Madison [9] have shed some light. The main reason for this lack of experiments is the complexity of generating ion pulses with the adequate characteristics to reproduce those of LIF. In the following sections, the authors propose two different techniques which can be suitable for that purpose and which can significantly contribute to the validation of current and new materials for fusion reactors.

## 2.1. Propagation of ion bursts through the chamber background gas/plasma: the two stream instability

It has already been indicated that the energy spectral distribution of LIF ions is of key importance in the behavior of the first wall and thus, on the design of the reaction chamber. According to their velocity, particles deposit their energy at different times and depths in the wall as well as inducing different atomistic effects on its lattice. Traditionally, the ion energy spectrum used to study those effects has always been the one at the origin of the burst and the interaction of those fusion ions with the surrounding gas/plasma systematically neglected. This fact is mostly because a classical description of the expected low pressure gas shows no moderation of the flying ions. However, those ions travel through an ionized gas, be it the one generated by the ablation of the target capsule or the one produced by the x-rays generated in the explosion. Since the expanding high energy ions with a broad band energy spectrum become monoenergetic locally in space and time (see figure 2), the well-known plasma instability the so-called two stream instability is expected to happen [10–13]. Then, a collective deceleration of the burst across the ambient plasmas due to the oblique ion–ion two stream instability and/or the ion–electron two stream instability can take place, resulting in collision-less shock which can significantly modify the final energy of the incident ions on the reactor wall.

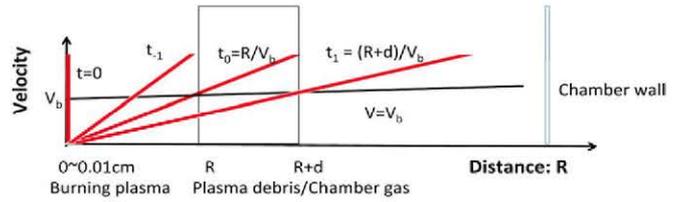

**Figure 2.** Schematic profiles of the interaction of the expanding burning plasma with the ambient plasmas (either the plasma created in the ablation of the pellet or the plasma created by the ionization of the residual chamber gas). The red lines are the phase space areas of the expansion. The burning plasma starts at $t = 0$ with velocity $V_b$ and interacts with the second plasma in the region $R$ and $R + d$. The burning plasma flies freely during the time interval $t - 1 = [0, R/V_b]$. From time $t0 = R/V_b$ to $t1 = (R + d)/V_b$ both plasmas interact. After the interaction, the expanding ion velocity decelerates and broadens.

As an example, the interaction of the ion bursts, in particular alpha particles, with the ablated plasma coming from the target shell is briefly outlined below (a more comprehensive paper is under construction). The implosion time for the reactor size target is typically 20 ns and the typical expansion velocity of the ablation plasma is $500\,\mathrm{km\,s^{-1}}$. Therefore, the radius of the ablated plasma, $R$, is approximately 1 cm at the burning time. The total ablated plasma will be of the order of 1 mg and the plasma mass density at this time will be $0.2\,\mathrm{mg\,cm^{-3}}$, namely ion/electron number density $n_0$ is about $10^{19}\,\mathrm{cm^{-3}}$. In the case of the alpha particles and according to Perkins simulations [4], they spread over 0.1–1 cm scale in 0.1–1 ns and the number density of alphas not decelerated in the core, $n_b$ will be $10^{18}\,\mathrm{cm^{-3}}$. Figure 2 shows the density and phase space distributions of the corona plasma and the alpha particles. This indicates that the alpha-particle expansion front is a monochromatic beam and will strongly interact with the background plasma through the ion–ion two stream instability. The growth rate of the instability is of the order of $\omega_{pi}(n_b/2n_o)^{1/3}$ where $\omega_{pi} \sim 10^{12}\,\mathrm{s^{-1}}$ and the e-folding distance of the instability propagating with the alpha particles is less than 0.1 mm. Therefore, it is expected that the alpha-particle expansion front will be mixed with the slower alpha particles quickly. The above electrostatic two stream instability will be quenched in that case, but the alpha-particle flow will be modulated laterally by the fluctuations and the generated wake fields (bow shocks) will decelerate the alpha particles. Furthermore, the electromagnetic instability may become effective in this process [14].

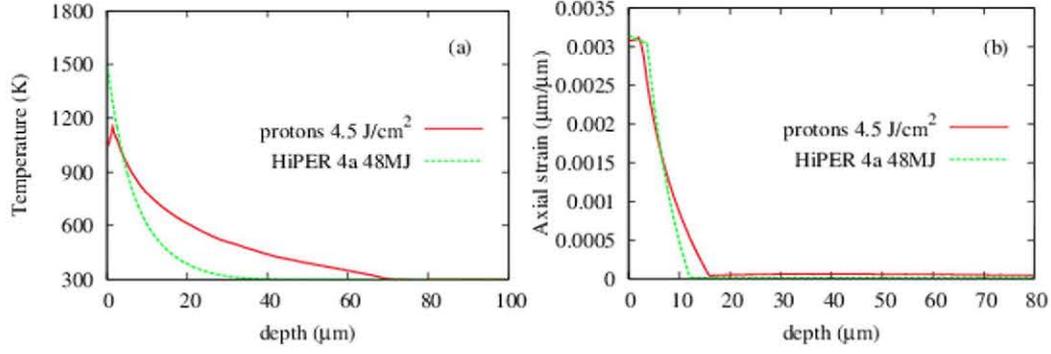

**Figure 3.** Response of a W slab to a laser proton beam of 4.5 J cm$^{-2}$ and the HiPER 4a 48 MJ scenario [5] at 300 K background temperatures. (*a*) The maximum temperature profile and (*b*) the permanent axial strain profile.

## 3. Laser-driven ions

The role of ultra-intense lasers on LIF might not be limited to their use as drivers to trigger the nuclear reaction but also as valuable tools to test materials under a fusion environment. In particular, these laser systems can be used as sources of ion pulses to investigate materials under irradiation and ion propagation through background gases/plasmas.

Among the different ion generation and acceleration mechanisms using ultra-intense lasers [15], the target normal sheath acceleration (TNSA) at the rear of laser targets happens to be the most suitable to simulate LIF ion bursts, i.e. high particle fluences, broad energy spectra and short durations [16, 17]. This ion acceleration mechanism and the resulting properties can be briefly explained as follows: electrons from a solid target irradiated by the ultra-intense laser are accelerated by the ponderomotive force to relativistic energies with mean free paths much larger than the thickness of the target itself. Hence, electrons accelerated towards the target, cross it and exit at the rear side creating a large sheath electrostatic field (TV m$^{-1}$) which in turn ionizes the atoms in that side of the target and accelerates them perpendicularly to the surface, forming a very directional, dense and short bunch. The duration of the ion bunch is directly related to the duration of the laser pulse and the motion of the electrons. Thus, the ion pulse can be as short as some ps and as long as some ns at its origin. As for the kinetic energy, the ions appear with a wide spectral energy because the electron spectrum is broadly spread due to electron–laser interaction and collisions. Thus, the typical energy spectrum of laser-accelerated particles can be approximated by a quasi-thermal distribution with a cutoff at a maximum energy related to the maximum ponderomotive force. In this way, the TNSA mechanism produces high flux and energetic ion beams very similar to those generated in LIF bursts. Throughout the literature there are numerous experimental examples and although in none of them researchers were aiming at generating ion spectra of a LIF reactor (traditionally, most of the research on laser ion acceleration is concentrated on generating very high monoenergetic beams), there are several references which back up the suitability of this approach. Here, the authors just refer to those which they consider more relevant for the species present in the fusion environment, namely, hydrogen [18], deuterium [19], carbon [20] and high Z atoms [21]. As for the generation of He pulses, which cannot be found in solid form, other acceleration mechanisms which favor/compete with TNSA also allow for the production and acceleration of ions [22].

Last but not least, it is important to mention that commercially available ultra-intense laser systems have the capability of performing those irradiation studies under repetition rates similar to those of LIF facilities, also providing the possibility to irradiate samples simultaneously with different ions or even with different radiation such as electrons or x-rays, thus, allowing for studies of synergistic effects.

### 3.1. Irradiation of materials: simulation and comparison of thermo-mechanical effects

The authors have already launched some experimental campaigns to use ultra-intense lasers to recreate fusion ion bursts and irradiate materials and the preliminary findings support our initial assumptions on the potentiality of this technique. Although results are still under discussion and more experiments will be needed before publication, a computational simulation of the effects of the two ion pulses on a tungsten sample is presented below. The thermo-mechanical response of a W slab under irradiation with a laser-driven proton pulse was calculated and compared to the one of a 48 MJ shock target [5] (simulations in [5] were done for a background temperature of 600 K; here are done for 300 K). A 'typical' TNSA proton pulse spectrum with the form $dN/dE = A\exp(-E/KT)$, with $A = 2 \times 10^{12}$, $KT \sim 3$ MeV and an angular full divergence of 20° is implemented in the simulations. In order to achieve an energy fluence on the sample of around 4.5 J cm$^{-2}$ similar to the one produced by the 48 MJ target on a 5 m radius chamber, the W slab was placed 3.13 cm away from the laser-driven proton source in the simulations. For the mentioned divergence of 20° and that distance, the irradiated area of the W slab corresponds to roughly 0.96 cm$^2$, yielding the appropriate energy fluence (this proton spectrum is typical of a hundred J energy laser, in order to achieve similar proton and energy fluences with a lower energy table-top laser, the distance between the proton source and the sample needs to be reduced). As in the case of the 48 MJ target, the spatial energy deposition profile was calculated using the FLUKA code (the temporal profile was estimated to be of a few hundred ps) and the thermo-mechanical response of the W was calculated using the CODE-ASTER. Figure 3(*a*) shows the temperature profiles of the W slab when the maximum temperature is reached in both cases, being a bit lower for laser protons due to the deposition of part of the energy at higher depths. If necessary, this difference can be corrected by decelerating the laser proton pulse using less energy or a moderator between the proton source and the irradiated sample. Figure 3(*b*) shows the permanent axial strain profile, deformation along the axial direction, after irradiation for the two scenarios. The deformation profiles are very similar in both scenarios with a deformation of 0.0031 $\mu$m/$\mu$m at the surface. Moreover, the permanent deformed volume is almost identical for both irradiations.

Having been carried out with a 'standard' laser proton beam, the similarities observed in these simulations are very enlightening, letting us conclude that laser-driven ion beams represent an excellent tool to validate materials.

### 3.2. Experimental simulations of ion burst propagation through the chamber background gas/plasma with laser-driven ions

The theoretical predictions described in section 2.1 on the propagation of ions through a background gas/plasma can be demonstrated by ultra-intense lasers using the following setup. When these lasers irradiate a solid metal film target, like a gold thin film target, the protons on the rear surface of the metal target (thin water layers are formed on the surfaces) are accelerated by the sheath field on the rear surface. These proton beams generated by TNSA have a very high current density and laminarity [15] and have been applied to the proton radiography to measure electromagnetic fields in plasmas [23]. If we set a gas plasma very close to the TNSA target and the gas plasma is irradiated by the TNSA proton beam with high current, the two stream instability will take place and their effects on the proton beam can be measured by taking the proton beam image after passing through the gas plasma. A typical setup of the simulation experiments is shown in figure 4. Typically, the proton beam energy is a few MeV and the total number of high energy protons is around $10^{11}$–$10^{13}$ per laser pulse. The proton current density can be higher than $10^{14}$ cm$^{-3}$ in the gas plasma. This current density is comparable to or higher than the alpha-particle density at the radius of 1 m in the LIF reactor chamber. When the gas density and size of the simulation experiments are about $10^{19}$ cm$^3$ and a few mm, respectively, the energy loss of protons by the two stream instability is significant. Initial experimental simulations with this setup have shown that the proton beam images at the radio chromic film (RCF) of figure 4 depend on the distance between the laser target and the helium gas jet. In order to observe those changes, a mesh was introduced between the laser target origin and the helium gas. When the distance is short (about 4 mm), the image of the mesh is strongly distorted, but not for the longer distances, like 8 mm. This indicates that the proton beam interacts with the gas jet collectively. Namely the stopping power scaling associated with the turbulence generated by the two stream instability is estimated to be (turbulent stopping power)/(beam energy): $1/L_{\text{STOP}} = 5.7 \times 10^2 n_a^{2/3} n_e^{-1/6} \times (\delta n_a/n_a)/V_a$ cm$^{-1}$, which yields 1% loss and 0.01 radian scattering, where beam density $n_a$ and electron density $n_e$ are in cm$^{-3}$, and $V_a$ is in cm s$^{-1}$. In future experiments, these effects will be further investigated to experimentally simulate the alpha-particle interaction with the background gas in a LIF reactor.

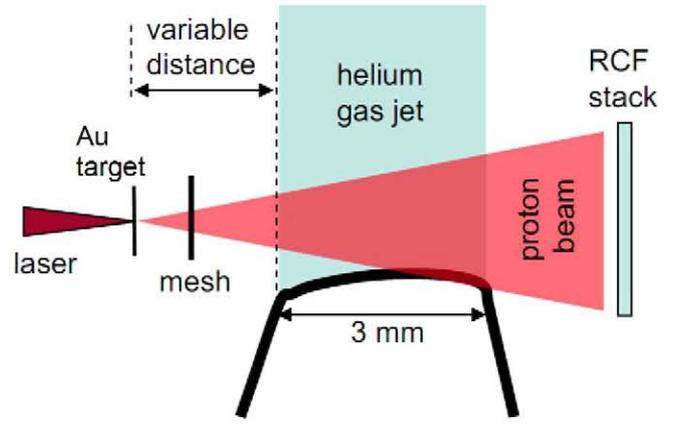

**Figure 4.** Experimental setup for the simulation of the two stream instability with laser-driven proton beams.

**Table 2.** Summary of HIPIB parameters for a stable plasma.

| | |
|---|---|
| $P_{\text{He}}$ (mbar) | $2.3 \times 10^{-2} - 3.5 \times 10^{-3}$ |
| $V_{\text{dc}}$ (V) | 600–700 |
| $I_{\text{dc}}$ (A) | 0.3–0.4 |
| $V_{\text{pulse}}$ (V) | 2–3 |
| $I_{\text{pulse}}$ (A) | 200–300 |
| Pulse width ($\mu$s) | 10–15 |
| Repetition (Hz) | 100–200 |

## 4. Ion pulses generated with HIPIB

As already mentioned, one of the main problems for material testing and qualification for fusion applications is the lack of existing facilities able to mimic the effects taking place in a nuclear fusion reactor environment. In particular, concerning inertial fusion, the major difficulty is to generate ion pulses with durations in the nano- to microsecond range delivering intensities of $\sim$10$^5$ MW m$^{-2}$, high flux parameter ($H$) [24] of $\sim$70 MJ m$^{-2}$ s$^{-1/2}$ due to the high thermal loads and ion implantation processes occurring in first wall materials [25]. Here we show the potential of a compact and versatile magnetron-based experimental setup to generate high intensity pulsed ion beams optimal for material validation. It is worthwhile to mention that the employment of HIPIB techniques for material modification is not new. Indeed, since the development of HIPIBs sources in the mid-1970s quite a lot of research has been carried out not only on the development of new technology for beam generation and transport to the target [26, 27] but also on practical applications such as short pulse ion implantation [28], surface modifications [29, 30]. A review of the existing HIPIBs facilities is reported in [27].

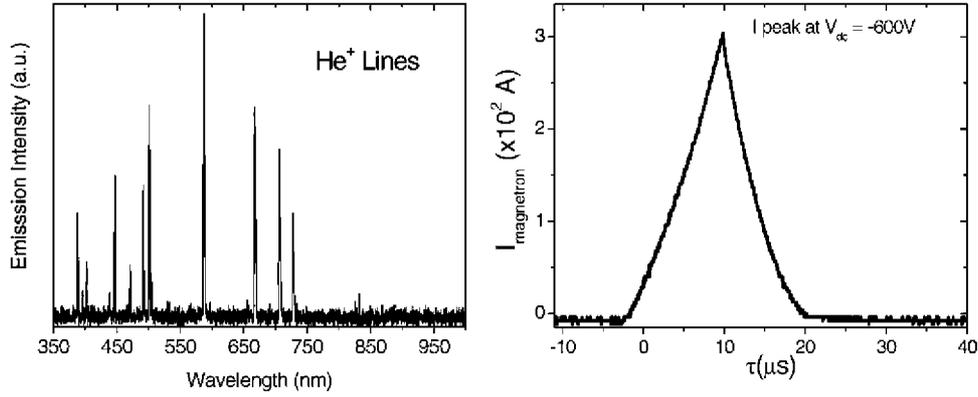

**Figure 5.** (*a*) Optical emission spectrum of the He plasma obtained at the optimal conditions. (*b*) Magnetron current as a function of pulse time duration.

The magnetron-based setup for the production of HIPIBs is located at the Institute of Microelectronics of Madrid (IMM/CSIC) [31]. It consists of a chamber with a base pressure in the $10^{-8}$ mbar range which is equipped with a 2 inch diameter magnetron sputtering source and a pulsed power supply, both developed and manufactured by Nano4energy SLNE [32]. This power supply is able to deliver peak power pulses of up to 0.48 MW (800 V and 600 A). The pulse is triggered by means of a 33220A Function/Arbitrary Waveform Generator Pulse which can generate variable-edge-time pulses up to 5 MHz with variable period, pulse width and amplitude. The parameters like voltage, repetition rate and pulse width can be tuned in order to modify the plasma/ion beam pulsed characteristics. A Rogowski coil delivered by PEM UK Ltd is used to characterize the high current pulses [33]. The chamber is connected to a gas inlet system which allows the introduction of diverse gases such as Ar, He and H, as convenient or a mixture of them. Gas pressure is monitored by a Pirani type gauge (Pfeiffer). The pumping speed can be adjusted using a gate valve and regulating the turbomolecular pump (TMP) speed. The main chamber is also equipped with a window which allows external diagnosis of the plasma. Optical emission spectroscopy (OES) is achieved by a CCD miniature spectrometer delivered by Ocean Optics.

Preliminary experiments to explore the setup capabilities to mimic nuclear fusion reactor radiation conditions were carried out using He gas. The requirements needed to achieve stable plasma are listed in table 2. A typical current time curve for the high power pulsed discharge giving rise to a heat flux parameter of $0.3\,\text{MJ m}^{-2}\,\text{s}^{-1/2}$ is shown in figure 5(*b*). This result is very promising even when the obtained $H$ value is about two orders of magnitude lower than required ($70\,\text{MJ m}^{-2}\,\text{s}^{-1/2}$). However, the $H$ value can be strongly increased by reducing the magnetron size and/or decreasing the pulse width. Work on this subject is being currently performed.

At gas pressures lower than $2.5 \times 10^{-2}$ mbar the magnetron current strongly decreases up to the mA range. At gas pressures higher than $3.5 \times 10^{-2}$ mbar the plasma stability vanishes because of the appearance of arcs. The optical emission spectroscopy was performed for the HIPIB discharges. The ion emission intensity is depicted in figure 5(*a*) (the optimal emission was obtained for a pulse of 11 µs and a frequency of 150 Hz). These data evidence the good plasma quality since all peaks present in the graphic correspond to ionization lines of He I.

## 5. Conclusions

Laser inertial fusion ion bursts are one of the main threats for direct drive dry-wall reactor chambers. Their broad and high energy spectra and short duration cause serious thermo-mechanical and atomistic effects on the walls which need to be assessed and understood to develop materials able to withstand the environments taking place in nuclear fusion reactors. Thus, any choice of material for the first wall requires a thorough investigation of operation performance under such radiation conditions. At present there are very few studies on the validation of materials under fusion ion bursts and more worrying, almost no facility in which these experiments can be carried out. This paper has presented two different techniques to overcome this lack, namely, ultra-intense laser systems and HIPIBs. By the TNSA mechanism, ultra-intense lasers are shown to generate suitable ion pulses which can be later used to irradiate materials and also investigate the propagation of such pulses under the background gases/plasmas present in a reaction chamber. The HIPIBs technique is very adequate for material testing. One advantage of using HIPIB for material testing is its flexibility under irradiation conditions just by tuning gas pressure, repetition frequency, pulse width, magnetron voltage, etc.

If exploited, these two techniques can significantly improve the research on materials for LIF applications, accelerating the design of novel first wall components and providing a more solid ground to the understanding of the propagation and moderation of fusion ions across background gases/plasmas.

## Acknowledgments

The authors thank the Spanish Ministry of Science and Innovation for economical support via the ACI-PROMOCIONA program 2009 (ACI2009-1040). Research by N Gordillo was supported by a PICATA postdoctoral fellowship of the Moncloa Campus of International Excellence (UCM-UPM).